\documentclass[letter,11pt]{article}
\pdfoutput=1

\usepackage{jheppub-mod}

\usepackage[T1]{fontenc}
\usepackage{amsthm}

\usepackage{listings}
\usepackage{mathrsfs}

\usepackage{xcolor}

\hypersetup{pdftitle={Inequalities of Holographic Entanglement of Purification from Bit Threads}, pdfauthor={Dong-Hui Du, Fu-Wen Shu}, citecolor=blue,linkcolor=blue,urlcolor=blue}

\newcommand{\be}{\begin{equation}}
\newcommand{\ee}{\end{equation}}

\usepackage{color}
\definecolor{purple}{rgb}{0.5,0,0.5}

\title{Inequalities of Holographic Entanglement of Purification from Bit Threads }

\author{Dong-Hui Du$^{1,2}$,}
\author{Fu-Wen Shu$^{1,2,3}$}
\author{and Kai-Xin Zhu$^{1,2}$}

\affiliation[1]{Department of Physics, Nanchang University, Nanchang, 330031, China}
\affiliation[2]{Center for Relativistic Astrophysics and High Energy Physics, Nanchang University, Nanchang, 330031, China}
\affiliation[3]{Center for Gravitation and Cosmology, Yangzhou University, Yangzhou, China}

\emailAdd{donghuiduchn@gmail.com}
\emailAdd{shufuwen@ncu.edu.cn}
\emailAdd{kaixinzhuabc@gmail.com}

\abstract{There are increasing evidences that quantum information theory has come to play a fundamental role in quantum gravity especially the holography. In this paper, we show some new  potential connections between holography and quantum information theory. Particularly, by utilizing the multiflow description of the holographic entanglement of purification (HEoP) defined in relative homology, we obtain several new inequalities of HEoP under a max multiflow configuration. Each inequality derived for HEoP has a corresponding inequality of the holographic entanglement entropy (HEE). This is further confirmed by geometric analysis. In addition, we conjecture that, based on flow considerations, each property of HEE that can be derived from bit threads may have a corresponding property for HEoP that can be derived from bit threads defined in relative homology.}

\begin{document}
\maketitle
\flushbottom

\section{Introduction}

The holographic essence of quantum gravity was elucidated in \cite{Holog1, Holog2}, which revealed the
duality between the quantum gravity theory in a $(d+1)$-dimensional space-time region and the quantum field theory on the $d$-dimensional boundary of this region. Specifically, in the AdS/CFT correspondence \cite{AdS/CFT1, AdS/CFT2, AdS/CFT3}, the entanglement entropy (the von Neumann entropy) for a spatial region $A$ on conformal boundary was shown to be given by the area of minimal homologous surface \cite{HEE1, HEE2}, i.e., the Ryu-Takayanagi (RT) formula
\begin{equation}\label{RTformula}
S(A)=\frac{\text{area}(m_{A})}{4G_{N}}\ ,
\end{equation}
where $m_{A}$ is the minimal surface in the bulk homologous to $A$. This reveals the deep connections between quantum entanglement and space-time geometry.

The entanglement entropy primely characterizes the quantum entanglement in pure bipartite states, and has many known properties. For example,
\begin{eqnarray}
&(1)&Araki-Lieb: |S(A)-S(B)|\leq S(AB), \label{pr1 EE}\\
&(2)&Subadditivity: S(AB)\leq S(A)+S(B), \label{pr2 EE}\\
&(3)&Strong\ Subadditivity\ 1: S(B)+S(ABC) \leq S(AB)+S(BC), \label{pr3 EE}\\
&(4)&Strong\ Subadditivity\ 2:S(A)+S(C) \leq S(AB)+S(BC). \label{pr4 EE}
\end{eqnarray}
It has been proved that the RT formula obeys all above properties of the entanglement entropy \cite{HEEPros1}. However, there is a property possessed by the holographic entanglement entropy (HEE) peculiarly \cite{ MMI1, MMI2, MMI3}, that is
\begin{eqnarray}
&(5)\ Monogamy: S(A)+S(B)+S(C)+S(ABC) \leq S(AB)+S(AC)+S(BC),\label{pr5 EE}
\end{eqnarray}
which is not obeyed by general quantum states. It gives a constraint on theories that potentially have a holographic duality. Alternatively, properties (\ref{pr1 EE})-(\ref{pr5 EE}) of HEE can be derived by the notion of bit thread \cite{BT0}, an alternative description of the HEE. See further works related to the bit threads in \cite{BT1, BT2, BT3, BT4, BT5, BT6, BT7, BT8, BT9, BT10, BT11, BT12}\footnote{Alternatively, in \cite{cEE} the authors interpret the RT surface as special Lagrangian cycles calibrated by the real part of the holomorphic one-form of a spacelike hypersurface.}.

Moreover, there is a quantity called entanglement of purification (EoP) \cite{EoP}, which is a measure of the classical correlations and quantum entanglement for mixed bipartite states. In \cite{HEOP1, HEOP2, HEoP1}, it has been conjectured that EoP is dual to the area of the minimal cross section on the entanglement wedge \cite{EW1, EW2, EW3} (and some related works in \cite{HEoP2, HEoP3, HEoP4, HEoP5, HEoP6, HEoP7, HEoP8, HEoP9, HEoP10, HEoP11, HEoP12, HEoP13, HEoP14, HEoP15, HEoP16, HEoP17, HEoP18, HEoP19, HEoP20, HEoP21, HEoP22, HEoP23, HEoP24, HEoP25, HEoP26, HEoP27, HEoP28, HEoP29, HEoP30, HEoP31, HEoP32, HEoP33, HEoP34, HEoP35}). For two non-overlapping spatial subregions $A$ and $B$ on the conformal boundary, we have
\begin{equation}\label{Ep=Ew}
E_{P}(A:B) =
\frac{\text{area}(\sigma^{min}_{AB})}{4G_N}\ ,
\end{equation}
where $\sigma^{min}_{AB}$ is the minimal cross section on the entanglement wedge. In \cite{BT8, BT10}, the bit-thread formulation of the holographic entanglement of purification (HEoP) was given, and many known properties of the HEoP were proved in this formulation. In this paper, however, we would like to go step further and to explore some new properties of HEoP by using bit threads.

According to the statement in \cite{BT1}, we can generalize the flow description of HEE into a meaning of relative homology, while the nesting property of flows goes through as before. Thus, we could carry these flow-based proofs of properties of HEE into homology case, which means we will obtain some corresponding properties in homology case. Remember that the flow description of HEoP in \cite{BT8, BT10} is based on the notion of relative homology. We will naturally obtain some corresponding properties for the HEoP,\footnote{Recently there is a work \cite{HEoP10}, where new inequalities of HEoP are obtained from HEE, based on the wormhole geometry description of HEoP. While we start from a flow viewpoint, and arrive at a similar conclusion.} as
\begin{eqnarray}
&(1)&|E_{P}(A:BC)-E_{P}(B:AC)| \leq E_{P}(AB:C),\label{pr1 EoP}\\
&(2)&E_{P}(AB:C)\leq E_{P}(A:BC)+E_{P}(B:AC),\label{pr2 EoP}\\
&(3)&E_{P}(B:ACD)+E_{P}(ABC:D)\leq E_{P}(AB:CD)+E_{P}(BC:AD),\label{pr3 EoP}\\
&(4)&E_{P}(A:BCD)+E_{P}(C:ABD)\leq E_{P}(AB:CD)+E_{P}(BC:AD),\label{pr4 EoP}\\
&(5)&E_{P}(A:BCD)+E_{P}(B:ACD)+E_{P}(C:ABD)+E_{P}(D:ABC)\nonumber\\
&&\leq E_{P}(AB:CD)+E_{P}(AC:BD)+E_{P}(BC:AD).\label{pr5 EoP}
\end{eqnarray}
These properties of HEoP obtained from bit threads do not follow from the known properties of EoP. In this paper, we will derive these inequalities by using multiflow description of HEoP defined in relative homology. A geometric analysis is also applied and the validity of these novel inequalities is further confirmed.

This paper is organized as follows: In section 2, we will briefly review the notion of bit threads. Then in section 3, we will have a discussion about generalized HEE defined in relative homology. In section 4.1, we give a multiflow description for HEoP in relative homology. Then in section 4.2, considering tripartite and quadripartite cases, we will derive out some properties of HEoP by using multiflows defined in relative homology, as inequalities (\ref{pr1 EoP})-(\ref{pr5 EoP}). We notice that each property for HEoP has a corresponding property of HEE in (\ref{pr1 EE})-(\ref{pr5 EE}). A concluding remark is given in the last section.

\section{Review of bit threads}

\subsection{Flow}

The bit threads \cite{BT0}, which are a set of integral curves of a divergenceless norm-bounded vector field $v$ with transverse density equal to $|v|$. The threads of a given vector field are oriented and locally parallel. Consider a manifold $M$ with a conformal boundary $\partial M$, where $A$ is a subregion on $\partial M$ and its complement is $\bar{A}:= \partial M \backslash A$. Define a flow $v_{A\bar{A}}$ with direction from $A$ to $\bar{A}$ on $M$, that is divergenceless and is norm bounded by $1/4G_{N}$:
\begin{align}
\label{flowdef}
\nabla_{\mu}v^{\mu}_{A\bar{A}} = 0\ , \quad |v_{A\bar{A}}| \leq \frac{1}{4G_{N}}\ .
\end{align}
As we set the direction for $v_{A\bar{A}}$ flowing from $A$ to $\bar{A}$, it means its flux given by $\int_A v_{A\bar{A}}$ is non-negative:
\begin{equation}\label{flux}
\int_A v_{A\bar{A}}:=\int_A\sqrt h\ n_{\mu}v^{\mu}_{A\bar{A}}\geq 0\ ,
\end{equation}
where $h$ is the determinant of the induced metric $h_{ij}$ on $A$ and $n_{\mu}$ is the (inward-pointing) unit normal vector. Then the entanglement entropy between $A$ and $\bar A$ is suggested to given by the maximum flux through $A$ among all flows:
\begin{equation}\label{maxflow}
S(A) = \max_{v_{A\bar{A}}}\int_A v_{A\bar{A}}.
\end{equation}
Equivalence between (\ref{maxflow}) and the RT formula (\ref{RTformula}) is guaranteed by the Riemannian MFMC theorem \cite{BT0}:
\begin{equation}
\max_{v_{A\bar{A}}}\int_A v_{A\bar{A}} = \min_{m\sim A}\frac{\text{area}(m)}{4G_{N}}\ .
\end{equation}
The left-hand side is a maximum of the flux over all flows $v$, while the right-hand side takes a minimum of the area over all surfaces $m$ homologous to $A$ (written as $m \sim A$). The flow interpretation of the holographic entanglement entropy, unlike the minimal surface captured by RT formula jumping under continuous deformations of region $A$ \cite{HTa, NTa, KKM, Hea}, varies continuously. And the subadditivity and the strong subadditivity inequalities of HEE can be proved by making use of the properties of flows \cite{BT0}.

\begin{figure}
\centering
\includegraphics[scale=0.950]{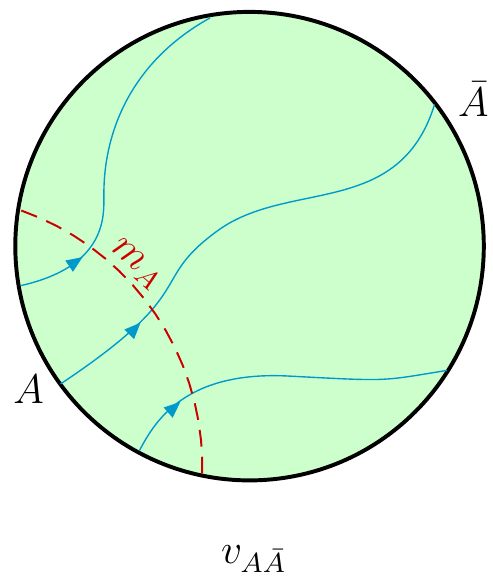}
\caption{The vector field $v_{A\bar{A}}$ defined on manifold $M$. The entanglement entropy $S_{A}$ is equal to the maximum flux from $A$ to $\bar{A}$, or equivalently the maximum number of threads connecting $A$ with $\bar{A}$.} \label{fig:flow1}
\end{figure}

\subsection{Threads}
In \cite{BT4}, the notion of bit threads was generalized. Instead of being oriented and locally parallel, threads are unoriented and can even intersect with others. The notion of $transverse\ density$ is replaced by $density$, defined at a given point on a manifold $M$ as the total length of the threads in a ball of radius $R$ centered on that point divided by the volume of the ball, where $R$ is chosen to be much larger than the Planck scale $G_{N}^{1/(d-1)}$ and much smaller than the curvature scale of $M$. In the classical or large-$N$ limit $G_{\rm N}\to0$, we can neglect the discretization error between the continuous flow $v$ and the discrete set of threads as the density of threads is large on the scale of $M$.

For region $A$ and its complement $\bar{A}$ on the boundary $\partial M$. Defining a vector field $v_{A\bar{A}}$, we can construct a thread configuration by choosing a set of integral curves with density $|v_{A\bar{A}}|$. The number of threads $N_{A\bar A}$ connecting $A$ to $\bar A$ is at least as large as the flux of $v_{A\bar{A}}$ on $A$:
\begin{equation}\label{Nfluxbound}
N_{A\bar A}\ge\int_A v_{A\bar{A}}\ .
\end{equation}
Generally, this inequality does not saturate as some of the integral curves may go from $\bar A$ to $A$ which have negative contributions to the flux but positive ones to $N_{A\bar A}$.

Consider a slab $R$ around $m$, where $R$ is much larger than the Planck length and much smaller than the curvature radius of $M$. The volume of this slab is $R\cdot area(m)$, the total length of the threads within the slab should be bounded above by $R\cdot area(m)/ 4G_{N}$. Moreover, any thread connecting $A$ to $\bar A$ must have a length within the slab at least $R$. Therefore, we have
\begin{equation}
N_{A\bar A}\le\frac{\text{area}(m)}{4G_{N}}\ .
\end{equation}
Particularly, for the minimal surface $m_{A}$, we have
\begin{equation}\label{Nbound}
N_{A\bar A}\le\frac{\text{area}(m_{A})}{4G_{N}}\ = S(A) \ .
\end{equation}
Combining formulas (\ref{Nfluxbound}) and (\ref{Nbound}), equality holds
\begin{equation}\label{Nboundtight}
\max N_{A\bar A}=\max_{v_{A\bar A}}\int_A v_{A\bar A} = S(A) \ .
\end{equation}
Thus, $S(A)$ is equal to the maximum number of threads connecting $A$ to $\bar{A}$ over all allowed thread configurations.

\subsection{Multiflow}\label{sec 2.3}

The \emph{multiflow} or \emph{multicommodity} is the terminology in the network context \cite{FKS, Sch}. It is a collection of flows that are compatible with each other, existing on the same geometry simultaneously. It was defined in Riemannian setting to prove the monogamy of mutual information (MMI) in \cite{BT4}. Consider a Riemannian manifold $M$ with boundary $\partial M$, and  let $A_1, \ldots, A_n$ be non-overlapping regions of $\partial M$, a \emph{multiflow} is then defined as a set of vector fields $\{ v_{ij} \}$ on $M$ satisfying the following conditions:
\begin{eqnarray}
&v_{ij}= -v_{ji},\label{antisym1}\\
&n_{\mu} v^{\mu}_{ij} =0\ \text{on}\ A_k\ (k \neq i,j),\label{noflux1}\\
&\nabla_{\mu} v^{\mu}_{ij}= 0,\label{divergenceless1}\\
&\sum_{i < j}^n |v_{ij}|\leq \frac{1}{4G_{N}} .\label{normbound1}
\end{eqnarray}
There are $n(n-1)/2$ independent vector fields for the given condition (\ref{antisym1}). Given condition (\ref{noflux1}), $v_{ij}$ is nonvanishing only on $A_{i}$ and $A_{j}$, by (\ref{divergenceless1}), their flux satisfy
\begin{equation}\
\int_{A_i}v_{ij}= -\int_{A_j}v_{ij}\ .
\end{equation}
Define a vector field
\begin{equation}
v_{i\bar{i}}:=\sum_{j\neq i}^n v_{ij}\ .
\end{equation}
The flux of flow $v_{i\bar{i}}$ should be bounded above by the entropy of $A_i$:
\begin{equation}\label{f-Si}
\int_{A_i}v_{i\bar{i}}\leq S(A_i)\ .
\end{equation}
The inequality will saturate for a max flow. Given $v_{ij}$ ($i<j$), we can choose a set of threads with density $|v_{ij}|$. From (\ref{Nfluxbound}), the number of threads connect $A_{i}$ to $A_{j}$ is at least the flux of $v_{ij}$:
\begin{equation}\label{ij bound}
N_{A_i A_j}\ge\int_{A_i}v_{ij}\ .
\end{equation}
Summing (\ref{ij bound}) over $j\neq i$ for fixed $i$, we have
\begin{equation}\label{N>flux}
\sum_{j\neq i}^n N_{A_i A_j} = N_{A_i \bar{A}_i} \ge \int_{A_i}v_{i\bar{i}} \ .
\end{equation}
On the other hand, (\ref{Nbound}) implies that the total number of threads emerging out of $A_{i}$ is bounded above by $S(A_{i})$:
\begin{equation}\label{Ni<EE}
\sum_{j\neq i}^n N_{A_i A_j} = N_{A_i \bar{A}_i} \leq S(A_{i}) \ .
\end{equation}
Therefore, both inequalities (\ref{N>flux}) and (\ref{Ni<EE}) saturate for a max flow with fixed $i$:
\begin{equation}\label{Ni=EE}
\sum_{j\neq i}^n N_{A_i A_j} = N_{A_i \bar{A}_i} = \int_{A_{i}} v_{i\bar{i}} = S(A_{i}) \ .
\end{equation}
Furthermore, the inequality (\ref{ij bound}) must be individually saturated:
\begin{equation}\label{ij=flow}
N_{A_i A_j} = \int_{A_i}v_{ij}\ .
\end{equation}
The above discussion focuses only on the case for a fixed $i$. Remarkably, it has been proved in \cite{BT4} that there exists a so-called \emph{max multiflow} $\{ v_{ij} \}$ saturating all $n$ bounds in (\ref{f-Si}) simultaneously. Or equivalently, there exists a so-called \emph{max thread configuration} in the language of threads, as the formula (\ref{Ni=EE}) holding for all $i$.

\section{Relative homology and generalized HEE}\label{Rel homo}

\begin{figure}
\centering
\includegraphics[scale=1.0500]{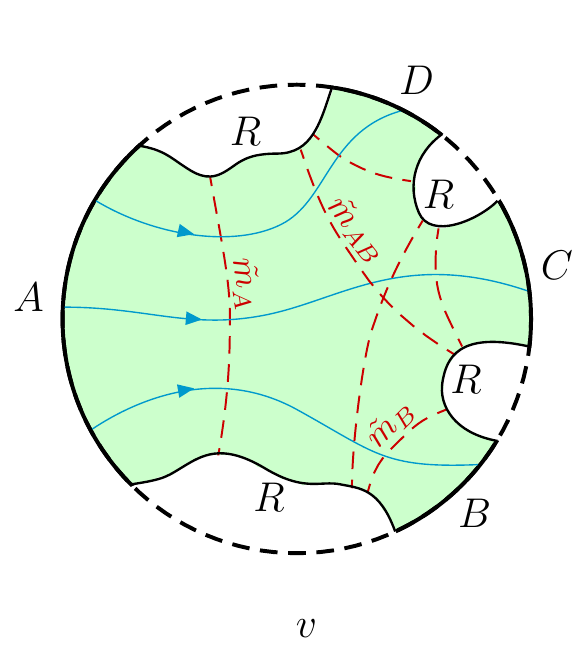}
\caption{The vector field $v$ is defined on $M'$, the region surrounded by $A\cup B\cup C\cup D\cup R$, as we have imposed a no-flux condition on surface $R$. And $\tilde{m}_{A}$, $\tilde{m}_{B}$ and $\tilde{m}_{AB}$ are the minimal surfaces homologous to regions $A$, $B$ and $AB$ relative to $R$, respectively. } \label{fig:gMFMC}
\end{figure}

The notion of relative homology was introduced to generalize the MFMC theorem in \cite{BT1}. To get an intuition for the generalized MFMC (gMFMC) theorem, we consider a manifold $M$ with a conformal boundary $\partial M$ and $A$ is a subregion on the boundary. Specifically, let $R$ be a bulk surface attached to the boundary, for instance as Figure \ref{fig:gMFMC}. For region $A$, we can define the surface $\tilde{m}$ homologous to $A$ relative to $R$ (written as $\tilde{m}\sim A\ rel\ R$), where surface $\tilde{m}$ is allowed to begin and end on $R$.\footnote{Here we use the notion $\tilde{m}$ defined on relative homology region $M'$, to distinguish from the surface $m$ defined on original manifold $M$.} On flow side, this corresponds to imposing a Neumann condition (no-flux condition) on $R$, thereby
\begin{equation}
\nabla_{\mu} v^{\mu}=0\ ,\  |v| \leq \frac{1}{4G_{N}}\ ,\  n_{\mu}v^{\mu}= 0 \text{ on }R.
\end{equation}
This means no flux through $R$. Therefore, the flow is restricted within region $M'$ with boundary $\partial M' =A\cup B\cup C\cup D\cup R$. Finally, we can arrive at the gMFMC theorem
\begin{equation}\label{gMFMC}
\max_{v: \atop n_{\mu} v^{\mu}\mid_{R} = 0}\int_{A} v\ = \min_{\tilde{m}\sim A \atop rel\ R\ on\ M}\frac{\text{area}(\tilde{m})}{4G_{N}}\ =\frac{\text{area}(\tilde{m}_{A})}{4G_{N}}\ ,
\end{equation}
where $\tilde{m}_{A}$ is the minimal surface homologous to $A$ relative to $R$ on region $M'$. The flow description of HEE in \cite{BT0} is specifically based on the homology relative to $R=\partial A$. As proposed in \cite{BT1}, we will have a generalized HEE
\begin{equation}\label{Sr}
\tilde{S}(A):=\max_{v: \atop n_{\mu} v^{\mu}\mid_{R} = 0}\int_{A} v\ .
\end{equation}
This notion originates from HEE, but defined in a meaning of general relative homology. We stress that, differing from HEE, this quantity has no clear quantum information meaning generally in holography. However, there is another quantity with the form (\ref{Sr}) arousing our interests in holography, i.e. HEoP. Similar to HEE, the HEoP can also be regarded as a special case in (\ref{Sr}) with the homology relative to $R=m_{AB}$ \cite{BT8, BT10}.

The formula (\ref{gMFMC}) establishes the equivalence between the flow objects and geometric objects in holography. We have learned that we could derive the properties of holographic objects from the properties of flows, such as nesting of flows. Remember the flow-based proofs of the Araki-Lieb (AL) inequality, the subadditivity (SA) and the strong subadditivity (SSA) for the HEE \cite{BT0}, also the multiflow-based proof of the monogamy of the mutual information (MMI) \cite{BT4, BT5}. Note that when the min cut is defined in homology relative to a bulk boundary $R$, the dual flow (or multiflow) is subjected to a no-flux boundary condition on $R$. The nesting property of flows in relative homology goes through as before \cite{BT1}. Thus these flow-based proofs for properties of HEE could be carried into the relative homology cases in similar manner, meanwhile what we need to do on geometric side is just replacing with the cuts defined in relative homology. Namely, based on the flow viewpoints, we argue that the properties of HEE will hold for quantity $\tilde{S}$ generally (including HEoP case), as
\begin{eqnarray}
&|\tilde{S}(A)-\tilde{S}(B)|\leq \tilde{S}(AB)\leq \tilde{S}(A)+\tilde{S}(B),\label{pr1 S}\\
&\tilde{S}(B)+\tilde{S}(ABC) \leq \tilde{S}(AB)+\tilde{S}(BC),\label{pr2 S}\\
&\tilde{S}(A)+\tilde{S}(C) \leq \tilde{S}(AB)+\tilde{S}(BC),\label{pr3 S}\\
&\tilde{S}(A)+\tilde{S}(B)+\tilde{S}(C)+\tilde{S}(ABC)\leq \tilde{S}_(AB)+\tilde{S}(AC)+\tilde{S}(BC),\label{pr4 S}
\end{eqnarray}
corresponding to the inequalities (\ref{pr1 EE})-(\ref{pr5 EE}) of HEE.\footnote{We emphasize that we only lists the inequalities of $\tilde{S}$ corresponding to some basic inequalities of HEE here. While our argument based on flow viewpoints allows us to reap more once we take other inequalities of HEE into consideration.} It seems that these inequalities could be intuitively obtained geometrically by making use of the minimality of relative homology surface. For example, as $AB\sim\tilde{m}_{A}\cup \tilde{m}_{B}\sim \tilde{m}_{AB}\ rel\ R$, due to the minimality of $\tilde{m}_{AB}$ among all surfaces homologous to $AB$ relative to $R$, we naturally obtain $|\tilde{m}_{AB}|\leq |\tilde{m}_{A}|+|\tilde{m}_{B}|$. As to $A\sim \tilde{m}_{A}\sim \tilde{m}_{AB}\cup \tilde{m}_{B}\ rel\ R$, due to the minimality of $\tilde{m}_{A}$ among all surfaces homologous to $A$ relative to $R$, we have $|\tilde{m}_{A}|\leq |\tilde{m}_{AB}|+|\tilde{m}_{B}|$. In the following, we will focus on the multiflow-based proofs of these inequalities for HEoP.

\section{Properties of HEoP derived from bit threads}

\subsection{Multiflow description of HEoP as relative homology}

The bit thread formulation of the HEoP has been given in \cite{BT8, BT10}, which is based on the gMFMC theorem. In this section, instead, we would like to apply the notion of multiflow defined in relative homology, to give a multiflow description for HEoP. Taking a manifold $M$ with non-overlapping subregions $A_1, \ldots, A_n$ on boundary $\partial M$. Consider the entanglement wedge $r_{A_{1}A_{2}...A_{n}}$ with boundary $\partial r_{A_{1}A_{2}...A_{n}}= \mathcal{A}\cup m_{A_{1}A_{2}...A_{n}}$, where $\mathcal{A}:= A_{1}\cup A_{2}\cup...\cup A_{n}$. We could define a multiflow $\{ v_{ij} \}$ on the entanglement wedge, subject to a Neumann boundary condition on $m_{A_{1}A_{2}...A_{n}}$. Namely,
\begin{eqnarray}
&v_{ij}= -v_{ji},\label{antisym2}\\
&n_{\mu} v^{\mu}_{ij} =0\ \text{on}\ m_{A_{1}A_{2}...A_{n}}, \label{Neumann2}\\
&n_{\mu} v^{\mu}_{ij} =0\ \text{on}\ A_k\ (k \neq i,j),\label{noflux2}\\
&\nabla_{\mu} v^{\mu}_{ij}= 0,\label{divergenceless2}\\
&\sum_{i < j}^n |v_{ij}|\leq \frac{1}{4G_{N}} .\label{normbound2}
\end{eqnarray}
This means no flux through $m_{A_{1}A_{2}...A_{n}}$ or no threads connecting to $m_{A_{1}A_{2}...A_{n}}$.  In this way, we restrict the multiflow inside geometry $r_{A_{1}A_{2}...A_{n}}$.  All following discussions will be based on such a multiflow configuration.

We set the direction of $v_{ij}$ as a flow from $A_{i}$ to $A_{j}$, which means the flux $\int_{A_{i}} v_{ij}$ out of $A_{i}$ (inward-pointing on $A_{i}$) is non-negative:
\begin{equation}\label{flux}
\int_{A_{i}} v_{ij}:=\int_{A_{i}}\sqrt h\ n_{\mu} v^{\mu}_{ij}\geq 0\ ,
\end{equation}
where $h$ is the determinant of the induced metric on $A$ and $n$ is chosen to be a (inward-pointing) unit normal vector. Given condition (\ref{Neumann2}) and (\ref{noflux2}) the fact that $v_{ij}$ is non-vanishing only on $A_{i}$ and $A_{j}$, combining (\ref{antisym2}) and (\ref{divergenceless2}), we get
\begin{equation}\label{interchange2}
\int_{A_{i}} v_{ij}= -\int_{A_{i}} v_{ji} = \int_{A_{j}} v_{ji} \geq 0 \ .
\end{equation}

Similarly as before, we can define a minimal cut $\tilde{m}_{A_{i}}$ on $r_{A_{1}A_{2}...A_{n}}$ homologous to region $A_{i}$ relative to $m_{A_{1}A_{2}...A_{n}}$, which is exactly the minimal cross section $\sigma^{min}_{A_{i}\bar{A}_{i}}$ where $\bar{A}_{i}:=\mathcal{A}\backslash A_{i}$. Here, the quantity $\tilde{S}$ is just $E_{P}$ as proposed in \cite{BT8, BT10}. The dual flow is defined as
\begin{equation}
v_{i\bar{i}}:=\sum_{j\neq i}^n v_{ij}\ .
\end{equation}
The flux of flow $v_{i\bar{i}}$ should be bounded above by the area of minimal cross section $\sigma^{min}_{A_{i}\bar{A}_{i}}$. Combining with the conjecture of $E_{P}=E_{W}$, we have
\begin{equation}\label{pf<cut}
\int_{A_i}v_{i\bar{i}}\leq \min_{\sigma_{A_{i}\bar{A}_{i}}\sim A_{i} \atop rel\ m_{A_{1}A_{2}...A_{n}}}\frac{\text{area}(\sigma_{A_{i}\bar{A}_{i}})}{4G_{N}}=E_{P}(A_{i}:\bar{A}_{i})\  .
\end{equation}
In addition, the number of threads connect $A_{i}$ to $A_{j}$ is at least the flux of $v_{ij}$:
\begin{equation}\label{ij bound2}
N_{A_i A_j}\ge\int_{A_i}v_{ij}\ .
\end{equation}
Let us consider a max multiflow (or equivalently max thread configuration), as introduced in section \ref{sec 2.3}, the inequality in (\ref{pf<cut}) saturates simultaneously for all $i$, thus
\begin{equation}\label{Nij EP}
\sum_{j\neq i}^n N_{A_i A_j} = N_{A_i \bar{A}_i} = \int_{A_{i}} v_{i\bar{i}} = E_{P}(A_{i}:\bar{A}_{i})  \ ,
\end{equation}
Furthermore, all inequalities (\ref{ij bound2}) individually saturated:
\begin{equation}\label{Nij flowij}
N_{A_i A_j} = \int_{A_i}v_{ij}\ .
\end{equation}

The bipartite case has been shown in \cite{BT8, BT10}. In what follows, we will study such a max multiflow configuration (or equivalently max thread configuration) for tripartite and quadripartite cases, to derive out some new inequalities of the HEoP. As we will show that for each property (considering only AL, SA, SSA, MMI) of HEE that can be derived from bit threads, there is a corresponding property for HEoP.

\begin{figure}
\centering
\includegraphics[scale=0.790]{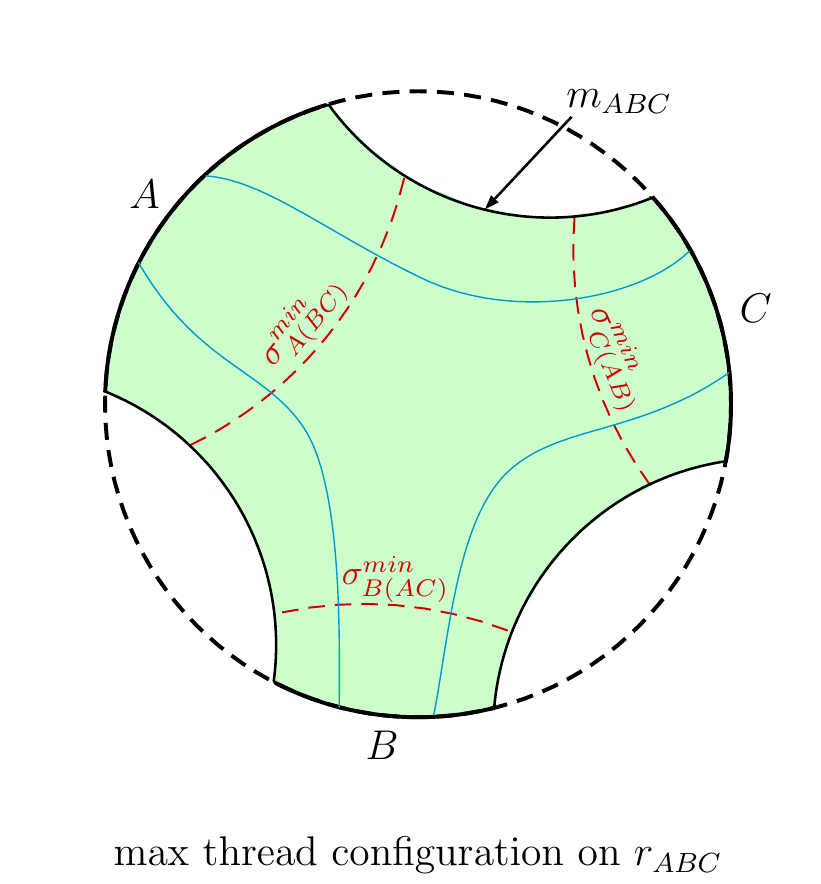}
\caption{The max thread configuration for tripartite case. Threads are unoriented and can intersect with others. The number of threads connecting to each single region reaches its maximum value.} \label{fig:flow2}
\end{figure}

\subsection{Properties of HEoP derived from bit threads}

\subsubsection{Tripartite case}

Consider a max multifow (or max thread configuration) for tripartite case with completely connected phase\footnote{Note that there are two other disconnected phases of region $r_{ABC}$. Subject to the no-flux condition on $m_{ABC}$, the disconnected phase means no-flow (or no-thread) passing between disconnected regions. Here, we just focus on completely connected case. Similar consideration is adopted for quadripartite case.}, as shown in figure \ref{fig:flow2}. By (\ref{Nij EP}) and (\ref{Nij flowij}), we have
\begin{align}\label{Ep12 N}
\begin{split}
&E_{P}(A:BC) = \int_{A} v_{A(BC)}= \int_{A} v_{AB}+\int_{A} v_{AC}= N_{AB}+ N_{AC},\\
&E_{P}(B:AC) = \int_{B} v_{B(AC)}= \int_{B} v_{BA}+\int_{B} v_{BC}= N_{BA}+ N_{BC},\\
&E_{P}(C:AB) = \int_{C} v_{C(AB)}= \int_{C} v_{CA}+\int_{C} v_{CB}= N_{CA}+ N_{CB}.
\end{split}
\end{align}
Then from (\ref{Ep12 N}) we obtain
\begin{align}
\begin{split}
E_{P}(A:BC)+E_{P}(B:AC)-E_{P}(AB:C)=2\int_{A} v_{AB}=2N_{AB},\\
E_{P}(AB:C)-(E_{P}(A:BC)-E_{P}(B:AC))=2\int_{B} v_{BC}=2N_{BC},\\
E_{P}(AB:C)-(E_{P}(B:AC)-E_{P}(A:BC))=2\int_{A} v_{AC}=2N_{AC},
\end{split}
\end{align}
where the relations (\ref{interchange2}) are used. Due to the non-negativity of the number of threads, we immediately get the inequalities
\begin{equation}\label{pr1 Ep}
|E_{P}(A:BC)-E_{P}(B:AC)| \leq E_{P}(AB:C)\leq E_{P}(A:BC)+E_{P}(B:AC).
\end{equation}
The first inequality can also be obtained from the second inequality by alternating labels A, B and C. The inequalities (\ref{pr1 Ep}) exactly correspond to the inequalities (\ref{pr1 S}) for HEoP case, i.e. the AL inequality and the SA. This property of HEoP has already been derived in \cite{BT8, HEoP10}. It does not follow by the known properties of EoP.

Moreover, this can also be intuitively obtained from geometric side by comparing with the cuts defined in relative homology, as shown in figure \ref{fig:flow2}. As $AB\sim \sigma^{min}_{A(BC)}\cup \sigma^{min}_{B(AC)} \sim \sigma^{min}_{(AB)C}\sim C\ rel\ m_{ABC}$, due to the minimality of $\sigma^{min}_{(AB)C}$ among all surfaces homologous to $AB$ relative to $m_{ABC}$, we naturally obtain the area relation $|\sigma^{min}_{(AB)C}|\leq |\sigma^{min}_{A(BC)}|+ |\sigma^{min}_{B(AC)}|$, that is, $E_{P}(AB:C)\leq E_{P}(A:BC)+E_{P}(B:AC)$ by HEoP conjecture. Similarly, as $A\sim \sigma^{min}_{A(BC)}\sim \sigma^{min}_{B(AC)}\cup \sigma^{min}_{(AB)C} \sim BC\ rel\ m_{ABC}$, due to the minimality of $\sigma^{min}_{A(BC)}$ among all surfaces homologous to $A$ relative to $m_{ABC}$, we get the inequality $|\sigma^{min}_{A(BC)}|\leq |\sigma^{min}_{B(AC)}|+ |\sigma^{min}_{(AB)C}|$, that is, $E_{P}(A:BC)\leq E_{P}(B:AC)+E_{P}(AB:C)$.

\subsubsection{Quadripartite case}

\begin{figure}
\centering
\includegraphics[scale=0.800]{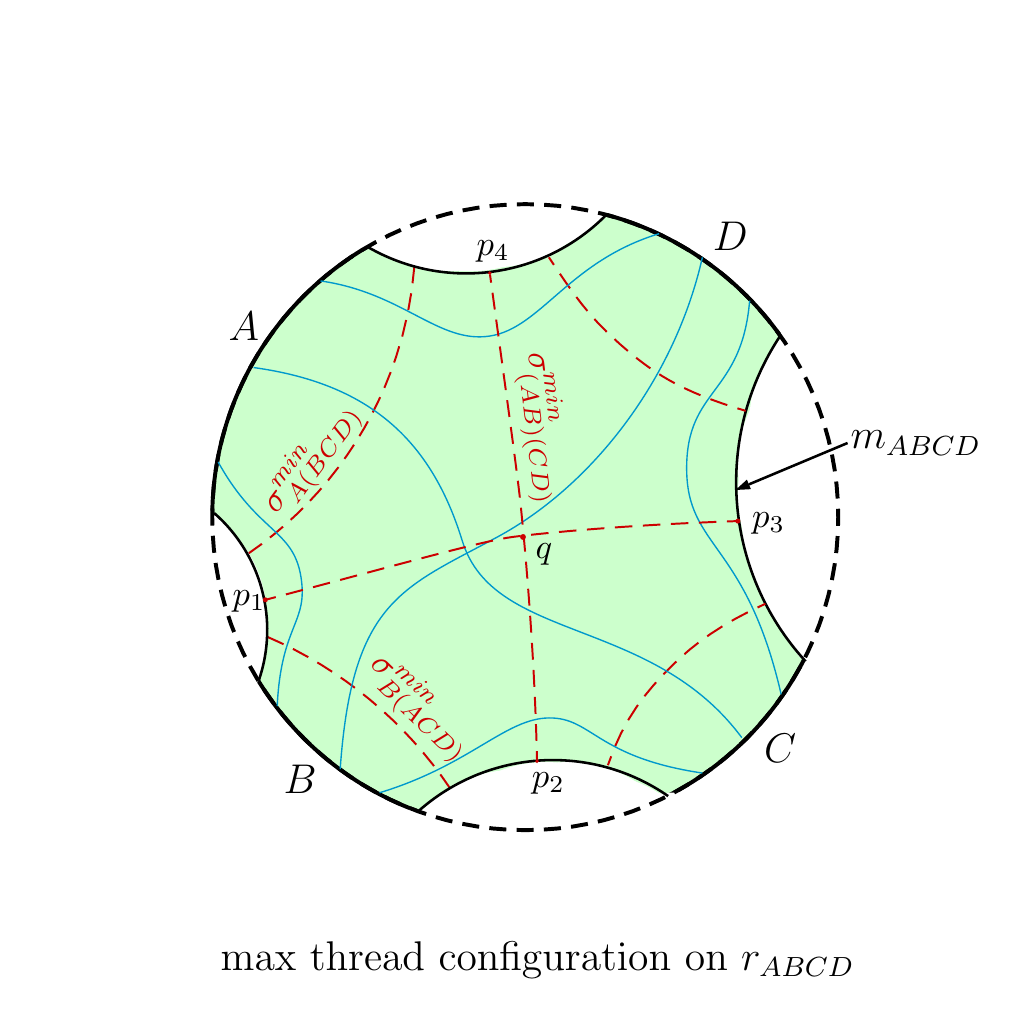}
\caption{The max thread configuration for quadripartite case. Threads are unoriented and can intersect with others. The number of threads connecting to each single region reaches its maximum value.} \label{fig:flow3}
\end{figure}

Consider a max multifow (or max thread configuration) for quadripartite case as sketched in figure \ref{fig:flow3}. From (\ref{Nij EP}) and (\ref{Nij flowij}), we have
\begin{align}\label{Ep13 N}
\begin{split}
E_{P}(A:BCD) = \int_{A} v_{A(BCD)}= \int_{A} v_{AB}+\int_{A} v_{AC}+\int_{A} v_{AD}= N_{AB}+ N_{AC}+ N_{AD},\\
E_{P}(B:ACD) = \int_{B} v_{B(ACD)}= \int_{B} v_{BA}+\int_{B} v_{BC}+\int_{B} v_{BD}= N_{BA}+ N_{BC}+ N_{BD},\\
E_{P}(C:ABD) = \int_{C} v_{C(ABD)}= \int_{C} v_{CA}+\int_{C} v_{CB}+\int_{C} v_{CD}= N_{CA}+ N_{CB}+ N_{CD},\\
E_{P}(D:ABC) = \int_{D} v_{D(ABC)}= \int_{D} v_{DA}+\int_{D} v_{DB}+\int_{D} v_{DC}= N_{DA}+ N_{DB}+ N_{DC}.
\end{split}
\end{align}
Noting that for a max multifow configuration, the flux through the union regions $AB$, $AC$ and $BC$ (or equivalently the number of threads emerging from these union regions) cannot reach its maximum value in general. Thus

\begin{align}\label{Ep22 N}
\begin{split}
E_{P}(AB:CD)\geq  \int_{AB} v_{(AB)(CD)}&= \int_{A} v_{AC}+\int_{A} v_{AD}+\int_{B} v_{BC}+\int_{B} v_{BD}\\
&= N_{AC}+ N_{AD}+N_{BC}+ N_{BD}\ ,\\
E_{P}(AC:BD)\geq  \int_{AC} v_{(AC)(BD)}&= \int_{A} v_{AB}+\int_{A} v_{AD}+\int_{C} v_{CB}+\int_{C} v_{CD}\\
&= N_{AB}+ N_{AD}+N_{CB}+ N_{CD},\\
E_{P}(BC:AD)\geq  \int_{BC} v_{(BC)(AD)}&= \int_{B} v_{BA}+\int_{B} v_{BD}+\int_{C} v_{CA}+\int_{C} v_{CD}\\
&= N_{BA}+ N_{BD}+N_{CA}+ N_{CD}.
\end{split}
\end{align}
Now from (\ref{Ep13 N}) and (\ref{Ep22 N}), and combining with relations (\ref{interchange2}), we obtain the inequality
\begin{align}\label{pr2 Ep}
\begin{split}
&E_{P}(AB:CD)+E_{P}(BC:AD)-E_{P}(B:ACD)-E_{P}(ABC:D)\\
&\geq \int_{AB} v_{(AB)(CD)}+\int_{BC} v_{(BC)(AD)}-\int_{B} v_{B(ACD)}-\int_{ABC} v_{(ABC)D}\\
&=2\int_{A} v_{AC}=2N_{AC}\geq 0,\\
\end{split}
\end{align}
and the inequality
\begin{align}\label{pr3 Ep}
\begin{split}
&E_{P}(AB:CD)+E_{P}(BC:AD)-E_{P}(A:BCD)-E_{P}(C:ABD)\\
&\geq \int_{AB} v_{(AB)(CD)}+\int_{BC} v_{(BC)(AD)}-\int_{A} v_{A(BCD)}-\int_{C} v_{C(ABD)}\\
&=2\int_{B} v_{BD}=2N_{BD}\geq 0.
\end{split}
\end{align}
Furthermore, we obtain that
\begin{align}\label{pr4 Ep}
\begin{split}
&E_{P}(AB:CD)+E_{P}(AC:BD)+E_{P}(BC:AD)\\
&-E_{P}(A:BCD)-E_{P}(B:ACD)-E_{P}(C:ABD)-E_{P}(D:ABC)\\
&\geq \int_{AB} v_{(AB)(CD)}+\int_{AC} v_{(AC)(BD)}+\int_{BC} v_{(BC)(AD)}\\
&-\int_{A} v_{A(BCD)}-\int_{B} v_{B(ACD)}-\int_{C} v_{C(ABD)}-\int_{D} v_{D(ABC)}=0.
\end{split}
\end{align}
Here we obtain the inequalities (\ref{pr2 Ep}), (\ref{pr3 Ep}) and (\ref{pr4 Ep}) by multiflows, which respectively corresponds to (\ref{pr2 S}), (\ref{pr3 S}) and (\ref{pr4 S}) for HEoP case. As far as we know, these properties of the HEoP are also new, and are out of the known properties of the EoP \cite{EoP}. It is worth to explore whether these properties are held by EoP for generic quantum states, or only valid for the holographic quantum states that have classical gravity duality.

Again, we can verify these properties from the geometric side. To do this, let us compare with these cuts defined in relative homology. First, we divide the union surface $\sigma^{min}_{(AB)(CD)}\cup \sigma^{min}_{(BC)(AD)}$ into four parts $(p_{1}q)$, $(p_{2}q)$, $(p_{3}q)$ and $(p_{4}q)$, as shown in figure \ref{fig:flow3}. Since $B\sim \sigma^{min}_{B(ACD)} \sim (p_{1}q)\cup(p_{2}q)$ and $D\sim \sigma^{min}_{(ABC)D} \sim (p_{3}q)\cup( p_{4}q)$ relative to $m_{ABCD}$, and considering the minimality of $\sigma^{min}_{B(ACD)}$ and $\sigma^{min}_{(ABC)D}$, we have
\begin{eqnarray}
|\sigma^{min}_{B(ACD)}|+|\sigma^{min}_{(ABC)D}|\leq |(p_{1}q)|+ |(p_{2}q)|+|(p_{3}q)| +|( p_{4}q)|=|\sigma^{min}_{(AB)(CD)}|+|\sigma^{min}_{(BC)(AD)}|,\nonumber
\end{eqnarray}
which is corresponding to the inequality (\ref{pr2 Ep}) by HEoP conjecture. Similar analysis leads to inequality (\ref{pr3 Ep}).

As to inequality (\ref{pr4 Ep}), we notice that $\sigma^{min}_{(AC)(BD)}= \min\{\sigma^{min}_{A(BCD)}\cup \sigma^{min}_{C(ABD)}, \sigma^{min}_{B(ACD)}\cup\sigma^{min}_{(ABC)D}\}$ for opposite-position regions. We then can choose that $|\sigma^{min}_{A(BCD)}|+|\sigma^{min}_{C(ABD)}| \leq
|\sigma^{min}_{B(ACD)}|+|\sigma^{min}_{(ABC)D}|$. In the end we obtain
\begin{eqnarray}
&&|\sigma^{min}_{(AB)(CD)}|+|\sigma^{min}_{(AC)(BD)}|+|\sigma^{min}_{(BC)(AD)}|-|\sigma^{min}_{A(BCD)}|-|\sigma^{min}_{B(ACD)}|-|\sigma^{min}_{C(ABD)}|-|\sigma^{min}_{(ABC)D}| \nonumber \\
&=&|\sigma^{min}_{(AB)(CD)}|+|\sigma^{min}_{(BC)(AD)}|-|\sigma^{min}_{B(ACD)}|-|\sigma^{min}_{(ABC)D}|.
\end{eqnarray}
Thus the area relation will reduce to the case for inequality (\ref{pr2 Ep}) or (\ref{pr3 Ep}), which we have proved in the previous paragraph. Finally, we finish the proofs from geometric side.

\section{Entropy cone for HEoP}

Section \ref{Rel homo} suggests a correspondence between the properties of the HEE and the ones of the HEoP. On the other hand, the work in \cite{BNOS}, which defines the holographic entropy cone of all entanglement entropy vectors for the holographic quantum states (see \cite{HEC1, HEC2, HEC3, HEC4, HEC5, HEC6} for more recent progress), reminds us to discuss the possibility to define an entropy cone for the holographic entanglement of purification. Particularly, we can likewise define an entropy cone for HEoP in terms of these inequalities (\ref{pr1 EoP})-(\ref{pr5 EoP}) derived in this paper. We notice that each inequality obtained for HEoP has the same form as the corresponding inequality of HEE. Thus the entropy cone of HEoP defined here will have the same structure as the one of HEE, except that it is defined in the entropy space of entanglement of purification.

For a tripartite state $\rho_{ABC}$, we choose $\{E_{P}(A:BC),\ E_{P}(B:AC),\ E_{P}(AB:C)\}$ as the basis of the entropy space. And the extreme ray of the HEoP cone is
\begin{equation}
(E_{P}(A:BC),\ E_{P}(B:AC),\ E_{P}(AB:C))=(1,\ 1,\ 0)
\end{equation}
up to the permutation symmetry. It reduces to the case of $\mathcal{C}_{2}$ of HEE when $\rho_{ABC}$ is pure, as $E_{P}(A:BC)=S(A),\ E_{P}(B:AC)=S(B),\ E_{P}(AB:C)=S(AB)$.

For a quadripartite state $\rho_{ABCD}$, after choosing the following basis, the extreme ray (up to the permutation symmetry) then is given by
\begin{align}
\begin{split}
&(E_{P}(A:BCD),\ E_{P}(B:ACD),\ E_{P}(C:ABD),\ E_{P}(D:ABC);\ E_{P}(AB:CD),\ \\
&E_{P}(AC:BD),\ E_{P}(BC:AD))=(1,\ 1,\ 1,\ 1;\ 2,\ 2,\ 2).
\end{split}
\end{align}
It reduces to the case of $\mathcal{C}_{3}$ of HEE when $\rho_{ABCD}$ is pure, as these HEoP sets are equal to HEE sets.

Furthermore, as stating in section \ref{Rel homo}, we can further argue that, for each inequality of HEE there always exists a corresponding inequality for HEoP, as it can be likewise proved by bit threads in the relative homology case.
If this is the case, it will help us define the entropy cone of HEoP for more partite cases by considering other inequalities of HEE. Moreover, besides these inequalities expected from HEE according to our argument, there are also other known inequalities for HEoP \cite{HEOP1, HEOP2, HEoP1}. Further study is needed to show whether these inequalities could together define a tighter entropy cone for HEoP to make constraints on the allowable holographic quantum states. We leave this study for the future work.

\section{Conclusion}

In this paper, we obtain some new properties (\ref{pr1 EoP})-(\ref{pr5 EoP}) of the HEoP that corresponds, respectively, to AL, SA, SSA and MMI by utilizing the multiflow description of the HEoP. Thus, starting from flow viewpoints, we arrive at a similar conclusion as in \cite{HEoP10}, and we give the multiflow-based derivations of these new inequalities. Note that there are several other quantum counterparts relating to the $E_{W}$, such as logarithmic negativity \cite{HEoP7, HEoP24}, odd entropy \cite{HEoP8}, reflected entropy \cite{HEoP21} and $R$-correlation \cite{HEoP25,HEoP28}. These new inequalities of HEoP should be also valid for other feasible counterparts, and make some constraints on these feasible candidates of $E_{W}$. Intuitively, these inequalities can also be obtained from geometric side by comparing with the cuts defined in relative homology for tripartite and quadripartite cases. But usually it could be more complicated and less obvious to find the inequalities of HEoP hiding in more partite cases from geometric side.

We can even go further, and conjecture that: For each property of HEE that can be derived from bit threads, there always exists a corresponding property for HEoP that can be obtained from bit threads defined in relative homology. The opposite is not true. We explain that all the flow-based proofs of the properties of HEE, can be carried into the relative homology cases in a similar manner, while on geometric side we just need to replace with the cuts defined in relative homology. Thus we could finally obtain some corresponding properties for HEoP. This is remarkable if it is true. As it allows one to explore much more constraints of $E_{W}$ by associating with the contents of holographic entropy cone. We leave the proof of this conjecture as a challenge for the future.

\section*{Acknowledgments}

We would like to thank Matt Headrick for useful discussion. This work was supported in part by the National Natural Science Foundation of China under grant number 11975116, 11665016, 11565019 and 11563006, and Jiangxi Science Foundation for Distinguished Young Scientists under grant number 20192BCB23007.

\end{document}